\documentclass{actasen}

\begin{document}

\setcounter{page}{001}

\Volume{2015}{XX}


\runheading{Comments on the spin down behavior of SGR J1745$-$2900}%

\title{Two predictions for the spin down behavior of magnetar SGR J1745$-$2900}

\footnotetext{$^{\dag}$ 
Received XXXX
\hspace*{5mm}$^{\bigtriangleup}$ tonghao@xao.ac.cn\\

\noindent 0275-1062/01/\$-see front matter $\copyright$ 2011 Elsevier
Science B. V. All rights reserved. 
\noindent PII: }

\enauthor{TONG Hao$^{\bigtriangleup}$ }{Xinjiang Astronomical Observatory, Chinese Academy of Sciences, Urumqi, 830011, China}

\abstract{The basic concepts of wind braking of magnetars, and its applications to various timing events of magnetars are outlined. The case of the Galactic centre magnetar SGR J1745$-$2900 is especially discussed. Two predictions for SGR J1745$-$2900 are made (the period derivative will first increase then decrease; and possible outbursts in the near future).  
}


\maketitle

\section{Introduction}

Magnetars are thought to be neutron stars whose radiations are powered by the star's magnetic energy (Duncan \& Thompson 1992).  Compared with normal rotation-powered pulsars, magnetars show various kinds of variabilities (Mereghetti et al. 2015), including short time scale bursts (and giant flares), and long lasting outbursts etc. At the same, magnetars also have many timing events (or timing irregularities): (1) timing noise and glitches (Dib \& Kaspi 2014); (2) decrease (or increase) of spin down rate by a factor of several or more (Archibald et al. 2015); (3) sudden spin down of the central star (i.e. anti-glitch; Archibald et al. 2013) etc. Furthermore, the radiative events and the timing events of magnetars are often closely related. 

One way to understand the timing events of magnetars is the wind braking model (Tong et al. 2013; see references therein for details and alternative models). The magnetic energy release may  first convert to a system of non-thermal particles. The radiation of these particles is responsible for the non-thermal radio, soft X-ray (even thermal X-rays by particle bombardment), hard X-ray radiations of magnetars. At the same time, the outflow of these particles will also carry away angular momentum of the central neutron star. The wind braking model can naturally account for the correlated radiative and timing events of magnetars. 

\section{Understanding timing events of magnetars in the wind braking model}

\subsection{Wind braking of magnetars and its applications}

Wind braking of pulsars and magnetars has been considered by many previous researchers (e.g. Harding et al. 1999; Xu \& Qiao 2001). Tong et al. (2013) pointed out that magnetar's spin down behavior may be dominated by the outflow of particle wind. The presence of additional magnetic energy is responsible for the strong particle wind. The magnetic dipole braking is incorrect even to the lowest order approximation in the case of magnetars. The full formulae of wind braking must be employed in order to understand the timing behaviors of magnetars. In the wind braking model, magnetars are neutron stars with strong multipole field (which is responsible for the burst, persistent emissions, and spin down of magnetars). A strong dipole field is no longer necessary. Then not only SGR 0418$+$5729, but also many other sources can have much lower dipole field. Both the particle wind and the X-ray emissions of magnetars are powered by the magnetic energy release. Therefore, it is natural that the radiative and timing events of magnetars are correlated (e.g. decreasing period derivative and X-ray luminosity during the outburst). 

Many peculiar timing behaviors of magnetars are understandable in the wind braking model. 
\begin{description}
\item[SGR 0418$+$5729]  Consider the effect of pulsar death, there exists another possibility of SGR 0418$+$5729. Since it is already near the death line, then its particle wind may have nearly stopped. The pulsar now is spun down mainly by the magnetic field perpendicular to the rotational axis. If the star has a small inclination angle, then its surface dipole field can also be as high as $10^{14} \,\rm G$. Therefore, the so called ``low magnetic field magnetar'' SGR 0418$+$5729 may be a normal magnetar (Tong \& Xu 2012).    

\item[Swift J1822.3$-$1606]  For magnetar Swift J1822.3$-$1606,
different timing observations found a different spin down rate.
This timing ambiguity may be solved if the period derivative is decreasing with time
(due to a decreasing particle luminosity).
Wind braking model predicted a long term averaged period derivative
$\dot{P} =1.9\times 10^{-14}$ (Tong \& Xu 2013, Section 2, last paragraph). Recent timing of this source
found a period derivative $\dot{P} =(2.1\pm 0.2) \times 10^{-14}$ (Scholz et al. 2014). 
It is consistent with the wind braking model.

\item[Anti-glitch] The magnetar 1E 2259$+$586 suffered a sudden spin down during an observational interval of two weeks (Archibald et al. 2013). The X-ray flux is two times higher than the quiescent state. The particle wind luminosity during this time interval may also be higher than the quiescent state. This enhanced particle wind will result in a net spin down of the central neutron star (Tong 2014a). 
Previous timing observations of SGR 1900$+$14, PSR J1846$-$0258, and SGR 1806$-$20 are consistent with the wind braking model of anti-glitch. If a stronger particle wind will result in net spin down (anti-glitch) of the central neutron star, a weaker particle wind may result in some spin up like event. This is may correspond to the spin down behavior of intermittent pulsars (Li et al. 2014). A stronger particle wind may account for the smaller braking index during glitches of the Crab pulsar (Kou \& Tong 2015).  

\end{description}

\subsection{Spin down behavior of the Galactic centre magnetar SGR J1745$-$2900}

SGR J1745$-$2900 is a radio emitting magnetar near the Galactic center. One holly grail of finding a radio pulsar orbiting the Milk Way supermassive black hole is partially fulfilled by this source. X-ray observations found an enhanced spin down rate (by a factor of two) when the X-ray luminosity keeps decreasing (Kaspi et al. 2014). This negative correlation between spin down rate and X-ray luminosity is hard to understand. Applying the wind braking model of magnetars (Tong et al. 2013) to the case of SGR J1745$-$2900, a changing geometry may explain this negative correlation. A smaller polar cap angle will result more particles to flow out. This corresponds to the enhanced spin down rate and decreasing X-ray luminosity simultaneously (Tong 2014b). If due to geometrical changes, one prediction for SGR J1745$-$2900 is: it should have a maximum spin down rate. Its spin down rate should first increase then decrease. Up to date observations (Coti Zelati et al. 2015) are consistent with the above geometrical explanations (after about one year, a further increase of spin down rate by a factor of two; Tong 2014b, section 3, paragraph 3 there). 

Besides the geometry, there may be physical changes in the magnetosphere of SGR J1745$-$2900. The magnetic field in the magnetosphere may be increasing (e.g. near the light cylinder). This results in the enhanced spin down rate. The increase of magnetic field means the magnetic energy is accumulating in the magnetosphere.  Once a critical point is reached, the magnetic energy can be release in a catastrophic way. If due to physical changes, a second prediction for SGR J1745$-$2900 is: it will enter into an outburst soon. 
This is similar to the previous case of 1E 1547.0$-$5408 (a decreasing X-ray luminosity accompanied by an increasing spin down rate, Camilo et al. 2008). It enters into outburst later (Mereghetti et al. 2009). If outburst is indeed observed in SGR J1745$-$2900 in the near future, it may indicate that the magnetar outburst (may also applicable to bursts and giant flares) is due to energy release in the magnetosphere. 
These two predictions may be related with each other. But the details are uncertain at present. 

\section{Conclusions}

The wind braking model can explain many timing events of magnetars. Future observations of SGR J1745$-$2900 will tell us whether wind braking is important in this source or not. 

\acknowledgements{H.Tong is supported Xinjiang Bairen project, West Light Foundation
of CAS (LHXZ201201), Qing Cu Hui of CAS, and 973 Program (2015CB857100).}

\end{document}